\documentclass[aip,showpacs]{revtex4-1}
\usepackage{graphicx}
\usepackage{amssymb}
\usepackage{amsmath,array}
\usepackage[letterpaper, portrait, margin=3cm]{geometry}
\begin{document}

\title{Density shift of Bose gas due to the Casimir effect and mean field potential } 



\author{G. M. Bhuiyan}
\email[]{gbhuiyan@du.ac.bd}
\affiliation{Department of Theoretical Physics, University of Dhaka, Dhaka-1000, Bangladesh}


\begin{abstract}
The shift of density of Bose gas due to the mean field potential (MFP) and the Casimir effect is systematically 
investigated in the $d$-dimensional configuration space from the point of thermodynamic consideration. We show that,
for $d=3$, the shift of density arises completely due to the Casimir effect and, the MFP remains totally ineffective
regardless of the state, condensate or non-condensate. But for dimension $d>3$ the MFP plays an active role 
in shifting the density of Bose gas along with the Casimir interaction. The sign of density shift becomes positive in the present case. So, the corresponding  critical temperature shift would be negative, because these two shifts are related as $\Delta n_{c}$/$n_{c}$ $\approx$ 
-$\Delta T_{c}$/$T_{c}$. It is important to note that, the MFP causes a shift of density for $d>3$ even when the Casimir effect is not there, and the sign of shift becomes negative then; consequently, the critical temperature shift
would be positive with MFP alone in this particular situation.
\end{abstract}

\maketitle 

\section{Introduction}
Since the demonstration by Einstein, a great many attempts have been made to understand the Bos-Einstein condensation(BEC) 
of the ideal Bose gas. Most of them are focused on the bulk properties including condensation through the thermodynamic route, occupation of the states and their fluctuations[1-5]. The effect of interparticle interaction on 
 the critical density and critical temperature of the Bose-Einstein condensation has attracted many theoretical efforts after the pioneering work by Lee and Yang[6], but it gets enhanced momentum after the observation of BEC in undercold atomic gases[7-9] in 1995.

The Casimir force due to the fluctuation of density of massless and massive bosons placed between two parallel slabs 
is well studied[10-25]. But we are not aware of any study focusing on the density shifts or critical temperature shift of Bose gas due to the Casimir interaction. In the present work, we study  the effect of the Casimir interaction on the bulk
density shift across the critical point of the Bose-Eienstein phase transition. We also show an interesting result that the mean field potential (MFP) does not affect in any way the density of Bose gas in three dimension, but does affect at higher dimensions. It is worth noting that, in some previous works[26,27] existing in literature the effect of dimensionality on BEC has been studied, but how the shift of density or critical density and crtical temperature are affected by the
dimensionality along with the MFP is yet to be explored. 

The critical density or critical temperature shift in BEC due to interactions or traps has been extensively explored
by many authors [27-35]. But results found for the amount of shift vary within certain extent, and become even 
contradictory in the sign of the shift. For example, in [28,30] it is predicted that the sign of $\Delta n_{c}$ 
is positive for repulsive interaction in dilute homogeneous Bose gas, whereas other works including measurement[29] and simulation studies[30,32] predict a negative sign. The corresponding sign for $\Delta T_{c}$ is just opposite, that is negative in the former cases and positive in latter ones, because the two shifts are related as $\Delta T_{c}$/$T_{c}$ $\rightarrow$ - $\Delta n_{c}$/$n_{c}$. Interestingly, the work by Giorgini et al. [25] shows that the density of
excited bosons increases due to repulsive interaction in an isotropic harmonic trap and, consequently, the critical
temperature goes down unlike homogeneous Bose gas. In the present study, we consider an ideal Bose gas system in a mean field potential and placed within two parallel walls. It is shown in an experiment [36] that the influence of interaction between particles on the Bose-Einstein condensation transition  temperature is only of few percent. This result fairly justifies the validity of our approach for a qualitative description of density shift of Bose gas.

The magnitude of the Casimir effect depends on the boundary conditions (bc) to be used. We note here that, the sums for the periodic, Neumann and Dirichlet boundary conditions starts at $-\infty, 0$ and $1$, respectively. However, the latter two bc can be expressed in terms of sums from $-\infty$ to $+\infty$. On the other hand, the operator relations for Neumann and Dirichlet bc are $\sum_{0}^{\infty}\rightarrow \frac{1}{2}(\sum_{-\infty}^{+\infty}+1)$ and $\sum_{1}^{\infty}\rightarrow \frac{1}{2}(\sum_{-\infty}^{+\infty} -1)$, respectively. The periodic bc is idealistic in the sense that the corresponding surface energy density becomes zero, while for others it is nonzero. We also note that the magnitude of Casimir interaction energy is the largest for periodic bc[37]. In this study, we use the Neumann boundary condition explicitly in the theory and then discuss the possible impact of other boundary conditions on the results when we feel it is relevant. 

This article is organized in the following way. Derivation of the grand canonical free energy for arbitrary dimension $d$ is made by using the one particle density of states,  in section 2. The general expression for the density shift is also presented and analysed for different dimensions in the same section. Section 3 is devoted to present the results and discussion. This paper is concluded in section 4.       
\section{Theoretical evaluation}
\subsection{The grand potential energy of the Bose gas in MFP and in $d$-dimensional configuration space}
The repulsive pair interaction between a pair of identical massive bosons can be described within the mean field theory as $a/V$,
where $a$ is a positive constant and $V$ is the volume containing the Bose gas. Let us now look at a particular boson
moving in the $N$ boson system in the mean field due to the rest of $(N-1)$ particles. Obviously, the average potential energy experienced by the tagged boson is $\frac{a}{V}(N-1)\approx \frac{a}{V}N$. The one boson Hamiltonian
will, therefore, be
\begin{equation}
H = \frac{p^{2}}{2 m} + a n
\end{equation}
where $p$ is the momentum of the boson and $m$ is its mass; $n$ denotes the number density of the bosons ($N/V$). 

Let us assume that, the bosons are enclosed in a $d$-dimensional volume $V^{(d)}=L^{d}$, $L$ being the edge of the rectangular box and
$L\rightarrow \infty$. The spacing between energy levels will therefore be very small, so the summation over
the states can be approximately replaced by integration. Therefore, the bulk density of states in the phase space with spatial $d$-dimension is
\begin{align}
\gamma(\epsilon)&=\frac{1}{(2 \pi \hbar)^{d}} \int d^{d}r \int d^{d}p \,\,\delta(\epsilon -\frac{p^{2}}{2\,m}-a\,n)\nonumber\\ &= V^{(d)}\,(\frac{m}{2 \pi\hbar})^{\frac{d}{2}}\, \frac{1}{\Gamma(\frac{d}{2})}\,(\epsilon-a\,n)^{\frac{d-2}{2}}
\end{align}
$\epsilon$ in equation (2) is the energy eigen value of the boson.
The density of states in the ($d$-1)-dimensional surface is 
\begin{align}
\gamma_{1}(\epsilon)= V^{(d-1)}\,(\frac{m}{2 \pi\hbar})^{\frac{d-1}{2}}\, \frac{1}{\Gamma(\frac{d-1}{2})}\,(\epsilon-a\,n)^{\frac{d-3}{2}}.
\end{align}

We now assume that the Bose gas in the mean field potential is placed within two parallel slabs such that the $d$-dimensional volume
$V^{(d)}=L^{d-1}\,D$, where $D$ is the separation distance of the slabs. 
As $D$ is finite, the spacing of the energy levels will be large along the $d-$th direction. So, the summation over energy levels along the $d-$th direction cannot be approximated by integration. The grand potential energy, using the Neumann boundary condition ( $k_{d}=\frac{\pi}{D}\,l;\, l= 0,1,2,3. ....$), may be expressed as
\begin{align}
\frac{\phi_{D}(T,\mu)}{k_{B}T}=&\sum_{l=0}^{\infty}\,\int_{0}^{\infty}\,ln(1-z\,\exp(-\frac{\beta \pi^{2} \hbar^{2}}{2mD^{2}}\,l^{2})\, \exp(-\beta\,\epsilon))\,\gamma_{1}(\epsilon)\,d\epsilon\nonumber\\=& -V^{(d-1)}\,\frac{1}{\lambda^{d-1}}\, \sum_{l=0}^{\infty}\sum_{r=1}^{\infty}\frac{z^{r}}{r^{\frac{d+1}{2}}}\nonumber\\ \times &\exp(-\frac{\pi \lambda^{2}l^{2}\,r}{4 D^{2}})\, \exp(-\beta a n r)\,\left\{\frac{1}{\Gamma(\frac{d-1}{2})}\Gamma(\frac{d-1}{2},-\beta a n r)\right\}
\end{align}
 where $z=\exp(\beta \mu)$, $\,\beta$ is inverse temperature times Boltzmann constant, $\mu$ the chemical potential, 
$\Gamma(\frac{d-1}{2},-\beta a n r)$  the lower incomplete gamma function and $\lambda=h/\sqrt{2 \pi m k_{B} T}$, $h$ being the Planck's constant. Now separating the $l=0$ term and using the Jacobi identity relation as in [18] we have
\begin{align}
\sum_{l=0}^{\infty}e^{-\pi \alpha l^{2}}&=1+\sum_{l=1}^{\infty}e^{-\pi \alpha l^{2}}\nonumber\\&=(\frac{1}{2\sqrt{\alpha}}+\frac{1}{2})+\frac{1}{\sqrt{\alpha}}\sum_{l=1}^{\infty} e^{-\pi\,l^{2}/\alpha}
\end{align}
Substituting equation(5) into (4) one can write,
\begin{align}
\frac{\phi_{D}(T,\mu)}{k_{B}T}&= -V^{(d-1)} \frac{1}{\lambda^{d-1}}\,\,\left[ \left(\frac{D}{\lambda}\right)\sum_{r=1}^{\infty}\frac{z^{r}}{r^{\frac{d+2}{2}}} +\frac{1}{2}\sum_{r=1}^{\infty}\frac{z^{r}}{r^{\frac{d+1}{2}}} 
   +2 \left(\frac{D}{\lambda}\right)\sum_{r=1}^{\infty}
\sum_{l=1}^{\infty} \frac{z^{r}}{r^{\frac{d+2}{2}}}\right. \nonumber \\
 & \times \,\left.e^{-4\pi(\frac{D}{\lambda})^{2}l^{2}/r}\right] 
 e^{-\beta a n r} \,\,
\left\{\frac{1}{\Gamma(\frac{d-1}{2})}\Gamma(\frac{d-1}{2},-\beta a n r)\right\} 
\end{align}
It appears from equation (6) that, the first and the third terms on right hand side are related to the bulk energy density,
whereas second term provides the surface energy density. According to Ref. 18 and 19 the third term is also recognized as
the Casimir interaction energy.

\subsection{Density shift}
From equation (6) it appears that the bulk energy density in $d$-dimension is
\begin{align}
\omega_{D}^{(d)}= -\frac{k_{B}T}{\lambda^{d}}\,\left[ \sum_{r=1}^{\infty}\frac{z^{r}}{r^{\frac{d+2}{2}}} 
   +2 \sum_{r=1}^{\infty}
\sum_{l=1}^{\infty} \frac{z^{r}}{r^{\frac{d+2}{2}}}
  e^{-4\pi(\frac{D}{\lambda})^{2}l^{2}/r}\right]\nonumber \\
 \times \, e^{-\beta a n r} \,\,
\left\{\frac{1}{\Gamma(\frac{d-1}{2})}\Gamma(\frac{d-1}{2},-\beta a n r)\right\} 
\end{align}
The bulk density of bosons may be obtained by differentiating the bulk energy density with respect to the 
chemical potential, $\mu$,
\begin{align}
n_{D}^{(d)}= - \frac{\partial \omega_{D}^{(d)}}{\partial \mu}=\frac{1}{\lambda^{d}}\,\left[ \sum_{r=1}^{\infty}\frac{z^{r}}{r^{\frac{d}{2}}} 
   +2 \sum_{r=1}^{\infty}
\sum_{l=1}^{\infty} \frac{z^{r}}{r^{\frac{d}{2}}}
  e^{-4\pi(\frac{D}{\lambda})^{2}l^{2}/r}\right]\nonumber \\
 \times \, e^{-\beta a n r} \,\,
\left\{\frac{1}{\Gamma(\frac{d-1}{2})}\Gamma(\frac{d-1}{2},-\beta a n r)\right\}
\end{align}

The lower incomplete gamma function is defined as
\begin{align}
\Gamma(q,x)=\Gamma(q)-\int_{0}^{x}e^{-t}t^{q-1}\,dt\hspace{1cm}\text{for}\,\, q>0
\end{align}
We can therefore write,
\begin{align}
\frac{1}{\Gamma(q)}\Gamma(q,x)=1-\frac{1}{\Gamma(q)}\int_{0}^{x}\,e^{-t}t^{q-1}\,dt
\end{align}
The above integration is not solvable exactly when $q$ takes on non integral values. But for integral values of $q$
it is analytically solvable. So, in this work, we shall solve the integration for integral values of $q$ only.

(i) CASE 1: $d=3$, $q=(d-1)/2=1$\\
For spatial dimension $d=3$, $q=(d-1)/2=1$. In this situation, the number density of Bose gas is,
\begin{align}
n_{D}^{(3)}= - \frac{\partial \omega_{D}^{(3)}}{\partial \mu}
=\frac{1}{\lambda^{3}}\,g_{\frac{3}{2}}(z) +\frac{2}{\lambda^{3}} \sum_{r=1}^{\infty}
\sum_{l=1}^{\infty} \frac{1}{r^{\frac{3}{2}}}
  e^{\beta \mu r-4\pi(\frac{D}{\lambda})^{2}l^{2}/r}.
\end{align}
The first term on the right hand side of equation(11) is just the free bulk number density, $n_{free}^{(3)}$, when the distance between two slabs $D=\infty$. So the shift in density
\begin{align}
\Delta n_{D}^{(3)}=n_{D}^{(3)}-n_{free}^{(3)}=\frac{2}{\lambda^{3}} \sum_{r=1}^{\infty}
\sum_{l=1}^{\infty} \frac{1}{r^{\frac{3}{2}}}
  e^{\beta \mu r-4\pi(\frac{D}{\lambda})^{2}l^{2}/r}.
\end{align}
(ii) CASE 2: $d=5$, $q=(d-1)/2=2$\\
For d=5, q = 2 and the bulk energy density is
\begin{align}
\omega_{D}^{(5)}=& - \frac{k_{B}T}{\lambda^{5}} 
\left[g_{\frac{7}{2}}(z)+2\sum_{r=1}^{\infty}\sum_{l=1}^{\infty}
\frac{1}{r^{\frac{7}{2}}}\, 
e^{\beta \mu r-4\pi(\frac{D}{\lambda})^{2}l^{2}/r}\right. \nonumber \\&-\left. \beta a n \left\{g_{\frac{5}{2}}(z)+2 \sum_{r=1}^{\infty}\sum_{l=1}^{\infty}
\frac{1}{r^{\frac{5}{2}}}\,e^{\beta \mu r-4\pi(\frac{D}{\lambda})^{2}l^{2}/r}\right\}\right].
\end{align}

The bulk particle density in five dimensional space,
\begin{align}
n_{D}^{(5)}=&  \frac{1}{\lambda^{5}} 
\left[g_{\frac{5}{2}}(z)+2\sum_{r=1}^{\infty}\sum_{l=1}^{\infty}
\frac{1}{r^{\frac{5}{2}}}\, 
e^{\beta \mu r-4\pi(\frac{D}{\lambda})^{2}l^{2}/r}\right. \nonumber \\&-\left. \beta a n \left\{g_{\frac{3}{2}}(z)+2 \sum_{r=1}^{\infty}\sum_{l=1}^{\infty}
\frac{1}{r^{\frac{3}{2}}}\,e^{\beta \mu r-4\pi(\frac{D}{\lambda})^{2}l^{2}/r}\right\}\right].
\end{align}
So, the amount of the density shift
\begin{align}
\Delta n_{D}^{(5)}=& \frac{1}{\lambda^{5}} 
\left[2\sum_{r=1}^{\infty}\sum_{l=1}^{\infty}
\frac{1}{r^{\frac{5}{2}}}\, 
e^{\beta \mu r-4\pi(\frac{D}{\lambda})^{2}l^{2}/r}\right. \nonumber \\&-\left. \beta a n \left\{g_{\frac{3}{2}}(z)+2 \sum_{r=1}^{\infty}\sum_{l=1}^{\infty}
\frac{1}{r^{\frac{3}{2}}}\,e^{\beta \mu r-4\pi(\frac{D}{\lambda})^{2}l^{2}/r}\right\}\right]
\end{align}

\subsection{Asymptotic Approximation when $\mu<0$:}
In the asymptotic limit i.e. for $D/\lambda \rightarrow \infty$ one can show as in Ref.18 that,
\begin{align}
\sum_{l=1}^{\infty}\sum_{r=1}^{\infty}\frac{1}{r^{\frac{d}{2}}}\,e^{\beta \mu r-4 \pi (\frac{Dl}{\lambda})^{2}/r} 
\leq
\frac{\zeta(\frac{d}{2})}{e^{4\sqrt{-\pi \beta \mu}\frac{D}{\lambda}}-1}\simeq\zeta(\frac{d}{2})\,e^{-4\sqrt{-\beta \pi \mu}\frac{D}{\lambda}}.
\end{align}
Equation (12) now reduces as
\begin{align}
\frac{\Delta n_{D}^{(3)}}{n_{c}^{(3)}}\approx 2\, e^{-4\sqrt{-\beta \pi \mu}\frac{D}{\lambda}} ,
\end{align}
where $n_{c}^{(d)}=\zeta(\frac{d}{2})/\lambda^{d}$.
Similarly, for $d=5$, equation (15) stands as
\begin{align}
\frac{\Delta n_{D}^{(5)}}{n_{c}^{(5)}}\approx 2\, e^{-4\sqrt{-\beta \pi \mu}\frac{D}{\lambda}}
 -\beta a n \left[\frac{
g_{\frac{3}{2}}(z)}{\zeta(\frac{5}{2})}+ 2\,\frac{\zeta(\frac{3}{2})}{\zeta(\frac{5}{2})}\,e^{-4\sqrt{-\beta \pi \mu}\frac{D}{\lambda}}\right].
\end{align}
\subsection{Density shift for $\mu \rightarrow 0$:}
For $d=3$, the density shift in the condensate (i.e. when $\mu=0$) can be obtained from equation (12), 
\begin{align}
\frac{\Delta n_{c}^{(3)}}{n_{c}^{(3)}}=\frac{2}{\zeta(\frac{3}{2})} \sum_{r=1}^{\infty}
\sum_{l=1}^{\infty} \frac{1}{r^{\frac{3}{2}}}
  e^{-4\pi(\frac{D}{\lambda})^{2}l^{2}/r}
\end{align}
Again, from equation (15) it follows for $d=5$ that,
\begin{align}
\frac{\Delta n_{c}^{(5)}}{n_{c}^{(5)}}=& \frac{1}{\zeta(\frac{5}{2})} 
\left[2\sum_{r=1}^{\infty}\sum_{l=1}^{\infty}
\frac{1}{r^{\frac{5}{2}}}\, 
e^{-4\pi(\frac{D}{\lambda})^{2}l^{2}/r}\right. \nonumber \\&-\left. \beta a n \left\{\zeta(\frac{3}{2})+2 \sum_{r=1}^{\infty}\sum_{l=1}^{\infty}
\frac{1}{r^{\frac{3}{2}}}\,e^{-4\pi(\frac{D}{\lambda})^{2}l^{2}/r}\right\}\right]
\end{align}
Now, the shift of critical temperature may be obtained, in principle, from the following relation[28],
\begin{align}
\frac{\Delta T_{c}}{T_{c}}\approx -\frac{2}{3}\frac{\Delta n_{c}}{n_{c}}.
\end{align}

\section{Results and Discussion}
Results for the density shift for the Bose gas in a MFP and placed between two parallel slabs is described in this section.
Equation (6) gives an exact account for the grand potential energy for Bose gas in MFP and for the given geometry 
in a $d$-dimensional configuration space. 
The energy apprantly depends on both the distance between slabs, $D$, and the mean field potential, $a n$. The 
first and the second terms on the right hand side of equation (6) are related to the bulk energy density and the surface  energy density,
respectively. The third term is related to the bulk energy density but arises due to the Casimir effect. If the MFP is 
switched off (i.e. set a=0) the third term purely accounts for the Casimir effect arising from the density fluctuation[10].
However, equation (7) shows the total bulk energy density in $d$-dimensional configurational space.The negative
of the derivative of the bulk energy density, $\omega_{D}^{(d)}$, with respective to the chemical potential, $\mu$,
yields the bulk number density of the Bose gas for the given situation. This is shown in equation (8). It is 
noticed that, equation (8) contains a lower incomplete gamma function $\Gamma(q,x)$. This is very complex to solve
when $q$ has a non-integral value (see Eqn. (10)). In order to make the integration simple and to analyze the effects
of the Casimir interaction and the MFP analytically, we consider the cases for $q=1 \,\text{and}\, 2$ only.
As $q=(d-1)/2$, values $1$ and $2$ for $q$ correspond to $d=3$ and $d=5$, respectively.

For $q=1$, $\Gamma(q,x)/\Gamma(q)=e^{-x}=e^{\beta a n r}$. So the bulk energy density becomes completely 
independent of the MFP. As a consequence, the density shift remains totally unaffected by the MFP. The most interesting thing happens when we go for higher dimension that is for $d > 3$. For $d=5$, $q=2$, and $\Gamma(q,x)/\Gamma(q)=e^{-x}(1+x)
=e^{\beta a n r}\,(1-\beta a n r)$. Equation (13), therefore, shows clearly that, for $d>3$ the bulk energy density depends 
not only on the Casimir effect alone but also on the MFP. As a result, for $d=5$, the shift in density of the Bose gas depends on both accordingly(see eq. (15)). Again for $q=$3 and 4, $\Gamma(q,x)/\Gamma(q)=e^{-x}(x^{2}+2 x+2)/2$ and $e^{-x}(x^{3}+3 x^{2}
+6 x+6)/6$, respectively. This is now crystal clear that the MFP contribution remains unvanished in all high dimensional cases, $d>3$.

The asymptotic approximation ($D/\lambda \rightarrow \infty$) of the density shift for $\mu < 0$ (that is in the
non-condensate state)is shown in 
equations (17) and (18) for $d=3$ and 5, respectively.  Equation (17) illustrates that, the shift in 
density is directly proprtional to $e^{-4\sqrt{-4\pi \mu}\frac{D}{\lambda}}$, which decays exponentially with 
increasing $D$, and finally vanishes at $D\rightarrow \infty$.
This is what we expected because when Casimir effect is zero density shift due to it must be zero.   As the $\Delta n_{D}^{(5)}/n_{c}^{(5)}$ for $d=5$ depends on both the Casimir effect and MFP, equation (18)
 demands a more careful analysis. It is seen that, if the MFP is switched off the ratio $\Delta n_{D}^{(5)}/n_{c}^{(5)}$ 
yields the same magnitude as that for $d=3$ and, varies with $D$ in a similar way. But if MFP remains on, the
ratio does not vanish for $D \rightarrow \infty$. The remainder is $\beta a n \,g_{\frac{3}{2}}(z)/\zeta(\frac{5}{2})$ 
which depends on the MFP. This means that, the MFP causes shift of the free bulk density in the non condensate state for $d=5$.

Equation (19) and equation (20) illustrate the shift of density for $d=3$ and 5, respectively, when $\mu \rightarrow 0$,
that is in the condensate. Here, for $d=3$, $\Delta n_{c}^{(3)}/n_{c}^{(3)}$ is solely governed by the Casimir effect alone. On the other hand, for $d=5$,
$\Delta n_{c}^{(5)}/n_{c}^{(5)}$ depends on both the Casimir interaction and the MFP; the former contribution is positive
and the latter contribution is negative. Now, if we accept the general relation between the critical density shift 
and the critical temperature shift [28,29], $\Delta n_{c}/n_{c} = -\frac{2}{3} \Delta T_{c}/T_{c}$, we can say that the 
critical temperature shift is negative for $d=3$ and could be positive or negative for $d=5$ depending on which dominates,
Casimir interaction or MFP. Here, it is worth noting that the sign of the Casimir interaction is geometry dependent [38], so, the sign of the corresponding density shift or critical temperature shift would alter accordingly.

\section{Conclusion}

The shift of density of Bose gas in MFP, and placed between two parallel slabs is systematically investigated from the
point of thermodynamic consideration. From the above results and discussions the following conclusions may be drawn.
For $d=3$, the density shift of Bose gas remains unaffected by the MFP regardless of the state condensate or non-condensate. The Casimir effect solely plays the role in shifting the density in this case. But for $d>3$ the MFP plays a significant role along with the Casimir interaction. Not only that, even in the absence of the Casimir effect the MFP provides a shift in density. To the best of my knowledge, this interesting feature of the Bose gas was not known before. Regarding the sign of the critical shift of density, it is found to be positive for $d=3$. Consequently, the corresonding sign of the shift in critical temperature of Bose gas would be negative (see equation (21)). This result contradicts with those obtained for repulsive interaction in the homogeneous gas. The attractive Casimir interaction in the present case clearly explains the cause of being negative sign in $\Delta T_{c}/T_{c}$. But for $d>3$, the critical temperature shift could be either positive or negative depending on which term Casimir effect or MFP dominates. If $\beta a n <1$, which is generally considered in the case of mean field, the sign of $\Delta T_{c}/T_{c}$ would be negative.  We finally say that, for $d>3$ and $D\rightarrow \infty$ the density shift appears only for the MFP and the sign of shift be negative. As a result, the sign of $\Delta T_{c}/T_{c}$ will be positive. This is analogous to the widely accepted results that for repulsive interaction in dilute homogeneous Bose gas the critical temperature 
shift would be positive (for $d=3$). For even integral or non-integral value of dimensionality $d$, the density shift may be evaluated by using an approximate analytic expression for the lower incomplete gamma function existing in the literature[39].

References\\\\
$[1]$ R. M. Ziff, G. E. Uhlenbeck and M. Kac, Phys. Reports 32, 169 (1977).\\
$[2]$ R. Beckmann, F. Karsch and D. E. Miller, Phys. Rev. Lett. 43, 1277(1979).\\
$[3]$ S. Fujita, T Kimura and Y. Zheng, Foundation of Phys. 21,1117(1991).\\
$[4]$ K Huang, Statistical Mechanics, John Wiley and Sons, Inc. New York 1963.\\
$[5]$ L. J. Landau and I. F. Wilde, Commun. Math. Phys. 70, 43 (1979).\\
$[6]$ T. D. Lee and C. N. Yang, Phys. Rev. 112, 1419(1958).\\ 
$[7]$ M. H. Anderson, J. R. Ensher, M. R. Matthews, C. E.Wieman and E. A.Cornell, Science 269, 198(1995).\\
$[8]$ C. C. Bradley, C. A. Sackett, J. J. Tollett and R. G. Hulet, Phys. Rev. Lett. 75, 1687 (1995).\\
$[9]$ K. B. Davis, M.-O. Mewes, M. R. Andrews, N. J. van Druten, D. S. Durfee, D. M. Kurn and W. Ketterle, 
           Phys. Rev.   Lett. 75, 3969 (1995).\\
$[10]$ H. B. G. Casimir, Proc. Kon. Nederl. Acad. Wetensch. 51, 793 (1948).\\
$[11]$ S. K. Lamoreaux, Phys. Rev. Lett. 78, 5 (1997).\\
$[12]$ Ariel Edery, J. Phys. A: Math. Gen. 39, 685 (2006).\\
$[13]$ U. Mohideen and Anushree Roy, Phys. Rev. Lett. 81, 4549 (1998).\\
$[14]$ M. Kardar, and R. Golestanian, Rev. Mod. Phys. 71, 1233 (1999).\\
$[15]$ G. Bressi, G. Carugno, R. Onofrio and G. Ruoso, Phys. Rev. Lett. 88, 041804-1 (2002).\\
$[16]$ P. A. Martin and J. Piasecki, Phys. Rev. E 68,016113(2003).\\
$[17]$ D. C. Roberts and Y. Pomeau, Phys. Rev. Lett. 95, 145303(2005).\\
$[18]$ P. A. Martin and V. A. Zagrebnov, Europhys. Lett. 73, 15 (2006).\\
$[19]$ M. Napiorkowski and J. Piasecki, Phys. Rev. E 84, 061105 (2011).\\
$[20]$ A. Gambassi and S. Dietrich, Europhys. Lett. 74, 754 (2006).\\
$[21]$ M. Napiorkowski and J. Piasecki, J. Stat. Phys. 147, 1145 (2012).
$[22]$ M. Napiorkowski, P. Jakubczyk and K. Nowak, J. Stat. Mech. 2013, P06015 (2013).\\
$[23]$ S. Biswas, Eur. Phys. J. D 42,109(2007).\\
$[24]$ T. Lin, G. Su, Q. A. Wang and J. Chen, Euro. Phys. Lett. 98, 40010 (2012).\\
$[25]$ S. Giorgini, L. P. Pitaevskii and S. Stringari, Phys. Rev. A 54, R4633 (1996).\\
$[26]$ M. Li, L. Chen, and C. Chen, Phys. Rev. A 59, 3109(1999).\\
$[27]$ L. Slasnich, J. Math. Phys. 41,8016(2000).\\
$[28]$ M. Holzmann, G. Baym, J-P Blaizo and F. Lalo$\ddot{e}$, Phys. Rev. Lett. 87, 120403 (2001).\\
$[29]$ V. A. Kasurnikov, N.V. Prokof'ev and B. V. Svistunov, Phys. Rev. Lett. 87,120402(2001).\\
$[30]$ P. Smith, R. L. D. Cambell, N. Tammuz and Z. Hadzibabic, Phys. Rev. Lett. 106, 250403 (2011).\\
$[31]$ G. Baym, J-P Blaizot, M. Holzmann, F Lalo$\ddot{e}$ and D. Vautherin, Phys. Tev. Lett. 83, 1703 (1999).\\
$[32]$ M. Holzmann and W. Krauth, Phys. Rev. Lett. 83, 2687 (1999).\\
$[33]$ P. Gr$\ddot{u}$ter, D. Ceperly and F. Lalo$\ddot{e}$, Phys. Rev. Lett. 79, 3549 (1997).\\
$[34]$ M. Wilkens, F. Illuminati and M. Kr$\ddot{a}$mer, J. Phys. B: At. Mol. Opt. Phys. 33, L779(2000).\\
$[35]$ H. T. C. Stoof, Phys. Rev. A 45, 8398 (1992).\\
$[36]$ J. R. Ensher, D. S. Jin, M. R. Matthews, C. E. Wieman and E. A. Cornell, Phys. Rev. Lett. 77, 4984 (1996).\\
$[37]$ E. B. Davies, Commun. Math. Phys. 28, 69(1972).\\
$[38]$ T. H. Boyer, Phys. Rev. 174, 1764(1968).\\
$[39]$ U. Blahak, Geosci. Model Dev. 3, 329 (2010).

\end{document}